\newcommand{\titulo}{Evolution of the Cosmological Horizons in a Universe with Countably Infinitely Many State Equations}
\newcommand{\tema}{Cosmology}

\documentclass[10pt,a4paper]{article}

\usepackage{amsmath}                    
\usepackage{amsfonts,amssymb,amsthm}    
\usepackage[english]{babel}                      
\usepackage[latin1]{inputenc}
\usepackage[breaklinks=true]{hyperref}  
\usepackage[T1]{fontenc}
\usepackage{appendix}
\usepackage[nottoc]{tocbibind}          

\usepackage{geometry}
\usepackage[width=.85\textwidth,font=small,labelfont=bf,labelsep=endash]{caption}

\usepackage{graphicx}
      \graphicspath{{}}

\usepackage{smartref}
      \addtoreflist{section}   
      \newcommand{\refconchap}[1]{\hyperref[#1]{\textcolor{\corlink}{\sectionref{#1}.}\ref*{#1}}}

\usepackage{fancyhdr}
\allowdisplaybreaks

    \newcommand{\corurl}{magenta}
    \newcommand{\corcite}{red}
    \newcommand{\corlink}{blue}
    \newcommand{\corfile}{black}

    \hypersetup{linktocpage,colorlinks,urlcolor=\corurl,citecolor=\corcite,linkcolor=\corlink,filecolor=\corfile, pdfnewwindow,pdftitle={\titulo},pdfauthor={Berta Margalef-Bentabol Juan Margalef-Bentabol},pdfsubject={\tema}}

    \newtheorem{theorem}{Theorem}[section]

    \theoremstyle{definition}
    \newtheorem*{remarks}{Remarks}

    \newtheorem{remark}[theorem]{Remark}
    
    \numberwithin{equation}{section}

    \makeatletter
      \newcommand{\raisemath}[1]{\mathpalette{\raisem@th{#1}}}
      \newcommand{\raisem@th}[3]{\raisebox{#1}{$#2#3$}}
    \makeatother
    \newcommand{\peqsub}[2]{#1_{\raisemath{-1pt}{\scriptscriptstyle #2}}}

\title{\titulo}
\author{Berta Margalef--Bentabol, Juan Margalef--Bentabol and Jordi Cepa}
\date{}

\begin{document}
    \pagestyle{fancy}
    \vspace*{2ex}

    \centerline{\textsf{\textbf{\Huge{Evolution of the Cosmological}}}}

    \mbox{}\vspace*{0.5ex}

    \centerline{\textsf{\textbf{\Huge{Horizons in a Universe with}}}}

    \mbox{}\vspace*{0.5ex}

    \centerline{\textsf{\textbf{\Huge{ Countably Infinitely Many}}}}

    \mbox{}\vspace*{0.5ex}

    \centerline{\textsf{\textbf{\Huge{State Equations}}}}

    \mbox{}\vspace*{2.5ex}

\centerline{
\begin{tabular}{ccccc}
    \textsf{\textbf{Berta Margalef--Bentabol}}{}\,$^1$ & \quad & \textsf{\textbf{Juan Margalef--Bentabol}}\,${}^{2,3}$ & \quad & \textsf{\textbf{Jordi Cepa}}\,${}^{1,4}$\\[1ex]
    \href{mailto:bmb@cca.iac.es}{bmb@cca.iac.es} & \quad & \href{mailto:juanmargalef@estumail.ucm.es}{juanmargalef@estumail.ucm.es} & \quad & \href{mailto:jcn@iac.es}{jcn@iac.es}
\end{tabular}}

\vspace*{2ex}

\begin{align*}
   {}^1 &\text{Departamento de Astrof\'isica, Universidad de la Laguna, E-38205 La Laguna, Tenerife, Spain}.\\[-0.6ex]
   {}^2 &\text{Facultad de Ciencias Matem\'aticas, Universidad Complutense de Madrid, E-28040 Madrid, Spain.}\\[-0.6ex]
   {}^3 &\text{Facultad de Ciencias F\'isicas, Universidad Complutense de Madrid, E-28040 Madrid, Spain.}\\[-0.6ex]
   {}^4 &\text{Instituto de Astrof\'isica de Canarias, E-38205 La Laguna, Tenerife, Spain.}\\
\end{align*}

\abstract{\noindent This paper is the second of two papers devoted to the study of the evolution of the cosmological horizons (particle and event horizons). Specifically, in this paper we consider the extremely general case of an accelerated universe with countably infinitely many constant state equations, and we obtain simple expressions in terms of their respective recession velocities that generalize the previous results for one and two state equations. We also provide a qualitative study of the values of the horizons and their velocities at the origin of the universe and at the far future, and we prove that these values only depend on one dominant state equation. Finally, we compare both horizons and determine when one is larger that the other.}
\mbox{}\vspace*{2ex}


\vspace*{3ex}

\begin{center}
  \begin{minipage}{95ex}
    \noindent This is an author-created, un-copyedited version of an article accepted for publication in Journal of Cosmology and Astroparticle Physics. IOP Publishing Ltd/SISSA Medialab srl is not responsible for any errors or omissions in this version of the manuscript or any version derived from it. The definitive publisher authenticated version is available online at:\vspace*{1ex}

    \hypersetup{urlcolor=blue}
    \centerline{\url{http://iopscience.iop.org/1475-7516/2013/02/015}}
    \hypersetup{urlcolor=\corurl}
  \end{minipage}
\end{center}

\setlength{\parindent}{0pt}      

\section{Introduction}
  In a Robertson-Walker universe, for any observer $A$ we can define two regions in the instantaneous three-dimensional space $t=t_0$. The first one is the region defined by the comoving points that have already been observed by $A$ (those comoving objects emitted some light in the past and it has already reached us), and the second one is its complement in the three-dimensional space, i.e. the region that cannot be observed by $A$ at a time $t_0$. The boundary between these two regions is the \textbf{particle horizon} at $t_0$, that defines the observable universe for $A$. Notice that the particle horizon takes into account only the past events with respect to $A$. Another horizon could be defined taking into account also the $A's$ future. This horizon is the \textbf{event horizon} and it is defined as the hyper-surface in space-time which divides all events into two classes, those that will be observable by $A$, and those that are forever outside $A's$ range of observation. This horizon determines a limit in the future observable universe \cite{Rindler}.\vspace{2ex}

  A deep study of the horizons has been made before in \cite{harrison} where only one state equation was considered. However, it is currently widely accepted the theory of a concordance Universe, i.e. that our present Universe is dominated by two state equations (cosmological constant and dust), one of them of negative pressure, that drives the universe into an accelerated expansion. In \cite{margalef_concordance} we obtained a generalization of \cite{harrison} for the case of a concordance universe. This generalization is physically acceptable for all times apart from those near to the Big Bang, where the radiation state equation is dominant over cosmological constant and matter equations. In this paper we are going to generalize all these results studying the most general case where countably many constant state equations are considered in an expanding universe.\footnote{Let us recall that countable stands for both finitely many or countably infinitely many, though in our case the finite case is trivially included in the infinite one.} Notice that we will deal with completely arbitrary state equations as long as the proportion factor remains constant (details are provided in the following sections). This condition is in fact not very restrictive, as only the phantom energy and some quintessences \cite{cepa2007cosmologia} are accepted to have non-constant state equations. Notice also that the particular case with any finite number of state equations is completely covered by our study.\vspace{2ex}

  This paper could be summarized as follows: in section two, we briefly introduce some basic cosmological concepts and establish the nomenclature to be used throughout the paper. In sections three and four, we define the formal concepts of particle and event horizons, and derive their integral expressions at any cosmological time. In section five we obtain the evolution of the horizons in a general universe with countably many constant state equations. In section six we compute the values of both horizons at the beginning of the universe and at the far future, we also study when one is larger than the other. Last section is devoted to gathering all the results obtained, discussion and conclusions. Throughout this paper, we use geometrical units where $c=G=1$.

\section{Cosmological Prerequisites}
  \subsection{Proper distance}
    In order to introduce the particle and event horizons, we need to define the \textbf{proper distance} $D_p$, which is the distance between two simultaneous events at a cosmological time $t_0$ measured by an inertial observer. Considering a homogenous and isotropic universe, we can write the Robertson-Walker metric, where $a(t)$ is the scale factor (albeit with units of longitude in our system) and $k$ the sign of the curvature:

    \begin{equation}\label{eq. RW}
        ds^2=-dt^2+a(t)^2\left(\frac{1}{1-k r^2}dr^2+r^2d\Omega^2\right)
    \end{equation}

    The definition of the proper distance and some physical considerations \cite[for more details see 2.1]{margalef_concordance} allow us to express the proper distance as a distance measured using light:

    \begin{equation}\label{def D_p}
        D_p(t)=a_0\int_t^{t_0}\frac{dt'}{a(t')}
    \end{equation}

    So to speak, the previous formula represents the distance covered by the light between two points of the space-time, but considering an expanding universe since the $a$ factor accounts for this expansion.

  \subsection{Hubble parameter}
    The \textbf{Hubble parameter} $H$ is defined as:

    \begin{equation}\label{def H(z)}
        H=\frac{1}{a}\frac{da}{dt}
    \end{equation}

    whose value at $t=t_0$ is the Hubble constant $H_0$. Notice in particular that for an expanding universe, $a$ is a strictly increasing function, then $H$ never vanishes. The product of $H$ times the proper distance, has dimensions of velocity and is known as the \textbf{recession velocity}:

    \begin{equation}\label{def recession}
        v_r\equiv H D_p
    \end{equation}

    which physically is the instantaneous velocity of an object at a distance $D_p$ with respect to an inertial observer.\vspace*{2ex}

    We are considering the universe as a perfect fluid with density $\rho$ and pressure $p$, but it can also be approximated as composed by non-interacting constituents of density $\rho_i$ and partial pressure $p_i$, where all together add up to $\rho$ and $p$ respectively. Each of this constituents is ruled by its own state equation $p_i=w_i\rho_i$. Now using second Friedmann equation \cite[chap.3]{kolb1990early} (where the dot stands for time derivation):

    \begin{equation}
        \dot{\rho}=-3H(\rho+p)=-3(1+w)\rho\frac{\dot{a}}{a}
    \end{equation}

    and considering the $i$-th state equation only, we can obtain the $i$-th density $\rho_i$ in terms of the redshift $z$:

    \begin{equation}\label{eq rho_i}
        \rho_i=\rho_{i0}(1+z)^{3\left(1+w_i\right)}
    \end{equation}

    Now we define the following time dependent magnitudes:

    \begin{itemize}
       \item[$\bullet$] Critical density as \ \ $\rho_c=\dfrac{3}{8\pi}H^2$
       \item[$\bullet$] Dimensionless energy density as \ \ $\Omega=\dfrac{\rho}{\rho_c}$
       \item[$\bullet$] Dimensionless $i$-th energy density as \ \ $\Omega_i=\dfrac{\rho_i}{\rho_c}$
    \end{itemize}

    All this quantities, when referred to the current time, are denoted with a zero subindex. If we now substitute in the definition of $\Omega_i$, the expressions of $\rho_c$, $\Omega_{i0}$ and $\rho_i$, we obtain:

    \begin{equation}\label{eq Omega_i}
        \Omega_i(z)=\Omega_{i0}H_0^2\frac{(1+z)^{3(1+w_i)}}{H(z)^2}
    \end{equation}

    Finally, adding all the $\Omega_i$ (where the index $i$ ranges over all possible state equations) leads to:

    \begin{equation}\label{H(z) sumatorio}
        H(z)=H_0\sqrt{\sum_{i=1}^\infty\frac{\Omega_{i0}}{\Omega}(1+z)^{n_i}}
    \end{equation}

    where $n_i=3(1+w_i)$ are real numbers that can depend on cosmic time or, equivalently, on $z$. However, in this work we will consider that they are all constant, which is indeed a very good approximation for the most commonly considered state equations: non-relativistic matter ($w=0$), radiation ($w=1/3$), cosmological constant ($w=-1$), or some quintessences ($-1<w<-1/3$).\vspace*{2ex}

    \begin{remarks}\mbox{}\renewcommand{\labelenumi}{\thesection.\arabic{enumi}}
      \begin{enumerate}
        \setcounter{enumi}{\value{theorem}}
        \item It is important to notice that the currently accepted theories admit only $-1\leq w_i\leq 1$ \cite{cepa2007cosmologia}, although in what follows we will not use this fact and we will consider the most general situation where $n_i$ are arbitrary constant.
        \item The case with finitely many state equation is included in the previous expression just by taking $\Omega_{i0}=0$ for every $i>K$, with $K$ the number of state equations.
        \item According to the first Friedmann equation:
            \[H^2=\frac{8\pi}{3}\rho-\frac{K}{a^2} \qquad \longleftrightarrow \qquad 1=\Omega-\frac{K}{a^2H^2}\]
            The curvature can be regarded as another component of the universe with density $\peqsub{\rho}{K}=\peqsub{\rho}{K0}(1+z)^2$ where $\peqsub{\rho}{K0}=-\frac{3K}{8\pi}$ (in particular we have $\peqsub{w}{K}=-1/3$ i.e. $\peqsub{p}{K}=-\peqsub{\rho}{K}/3$). The definition of $\Omega_i$ applied to this particular case tell us that$\peqsub{\Omega}{K}=-\frac{K(1+z)^2}{H^2}=-\frac{K}{a^2H^2}$ and therefore the Friedmann equation reads $1=\Omega+\peqsub{\Omega}{K}\equiv\peqsub{\Omega}{T}$. Hence, by including the curvature as a constant state equation we obtain that the addition of all the $\Omega_i$ (including $\peqsub{\Omega}{K}$) add up to $1$ and so we can take without loss of generality $\Omega=1$.\label{remark chap curvature}
        \item We will assume the following hypothesis on $H$ (where $S_n(z)$ is the finite summation up to the term $n$ and $F$ the infinite summation):
            \begin{center}
               The derivatives of $S_n$ converge uniformly to a function $G$ on $I=[-1,\infty)$, that means that:
               \[\sup_{z\in I}\left|S'_n(z)-G(z)\right|\overset{n\rightarrow \infty}{\xrightarrow{\hspace*{7ex}}}0\]
            \end{center}
        \item We can easily weaken this condition by asking the uniform convergence on $[-1,b]$ for every $b>-1$ (notice that this does not imply the uniform convergence on $[-1,\infty)$), but we have to be careful when defining the function $G$.
        \item This hypothesis, together with the fact that the series converges pointwise at $z=0$, implies two important facts. First, it assures us that the function $H$ is well defined on $[-1,\infty)$, and in the second place it ensures its differentiability and that its derivative can be obtained differentiating term by term \cite[9.41]{ponnusamy2011foundations}:
             \[F'(z)=\sum_{i=1}^\infty n_i\Omega_{i0}(1+z)^{n_i-1}\]
        \item Notice that this hypothesis is obviously attained if we consider finitely many state equations.
        \setcounter{theorem}{\value{enumi}}
      \end{enumerate}
    \end{remarks}

\section{Particle Horizon}
  \subsection{At the present cosmic time}\label{subsec Hp_0}
    As the age of the Universe and the light velocity have finite values, there exists a particle horizon $H_p$, that represents the longest distance from which we can retrieve information from the past, so it defines the past observable universe. Then the particle horizon $H_p$ at the current moment is given by the proper distance measured by the light coming from $t=0$ (the origin of the universe, where a hot big bang is assumed, compatible with the dominant equation of state) to $t_0$:

    \begin{equation}\label{def H_p}
        H_p=\lim_{t\rightarrow  0}D_p(t)
    \end{equation}

    It is a well known fact that the scale factor $a$ is related with the redshift $z$ as:

    \begin{equation}\label{def redshift}
        1+z=\frac{a_0}{a}
    \end{equation}

    Applying the previous equation, its derivative, the definitions of $H$ and $H_p$, it can be proved \cite[for more details see 3.1]{margalef_concordance} that

    \begin{equation}\label{H_p0}
        H_p=\frac{1}{H_0}\int_0^\infty\frac{dy}{\sqrt{\sum\Omega_{i0}(1+y)^{n_i}}}
    \end{equation}

  \subsection{At any cosmic time}\label{subsec Hp(z)}
    We have so far obtained the particle horizon for the current cosmic time (corresponding to $z=0$). If we want to know the expressions of this horizon at any cosmic time, or equivalently at any other $z'$, we have to replace in \eqref{H_p0} the constant parameters $H_0$ and $\Omega_{i0}$ (corresponding to $z=0$) with the parameters corresponding to $z'$ that we denote $H(z')$ and $\Omega_{i}(z')$. Where we recall from equation \eqref{eq Omega_i} that $\Omega_{i}(z')$ is $i$-th dimensionless energy density given by:

    \begin{equation}
        \Omega_i(z')=\Omega_{i0}\frac{H_0^2}{H(z')^2}(1+z')^{n_i}
    \end{equation}

    where $n_i=3(1+w_i)$. Therefore, if we replace all these constants in equation \eqref{H_p0} and simplify the result, we obtain the particle horizon for any redshift $z$ (where the prime is omitted in order to simplify the notation):

    \begin{equation}\label{H_p(z) general}
        H_p(z)=\frac{1}{H_0}\int_0^\infty\frac{dy}{\sqrt{\sum \Omega_{i0} (1+z)^{n_i}(1+y)^{n_i}}}
    \end{equation}

   That expression can be rewritten in a more suitable way by making the change of variable given by $x=(1+z)(1+y)-1$, then considering the definition of $H$ given by eq. \eqref{H(z) sumatorio} we obtain:

   \begin{equation}\label{H_p(z) final}
        H_p(z)=\frac{1}{1+z}\int_z^\infty\frac{dx}{H(x)}
    \end{equation}

    To obtain a characterization of the convergence of this last integral we need the following result of basic calculus \cite[sect.7.1]{ponnusamy2011foundations}:

    \begin{theorem}\mbox{}\\
     Let $I=[a,b)$ where $b>a$ and possibly $b=\infty$, and let $f,g$ be positive and continuous functions in $I$ such that
     \[ \lim_{x\rightarrow b^{-}}\frac{f(x)}{g(x)}=L>0 \]
     then $\int_a^b f(x)dx$ and $\int_a^b g(x)dx$ either both converge or both diverge.\vspace*{2ex}
    \end{theorem}

    Obviously, we have analogous results for $I=(a,b]$. In what follows, we will denote $N$ the highest $n_i$ which appears in $H(z)$  and $m$ the lowest one (we allow them to be $\pm\infty$ if the set $\{n_i\}_{i\in\mathbb{N}}$ is not upper/lower bounded).\vspace*{2ex}

    If we take $f(x)=H(x)$ and $g(x)=(1+x)^{N/2}$, then its ratio tends to $H_0\sqrt{\Omega_{N0}}>0$ when $x\rightarrow\infty$, so the integrals of $1$ over these functions behave in the same way. It can be easily proven \cite{ponnusamy2011foundations} that for any $z>-1$, the $p$-integral of $1/g$ converges if and only if $N/2>1$. Then we have the following characterization:\vspace*{2ex}

    \centerline{\fbox{$H_p(z)$ exists for $z>-1$ if and only if $N>2$}}
    \mbox{}

    The limit cases $z=-1$ and $z=\infty$ are studied in section \ref{sect values}.

\section{Event Horizon}
  \subsection{At the present cosmic time}
    The event horizon represents the barrier between the future events that can be observed, and those that cannot. It sets up a limit in the future observable universe, since in the future the observer will be able to obtain information only from events which happen inside their event horizon. According to its definition the event horizon can be expressed as:

    \begin{equation}\label{def H_e}
        H_e=\lim_{t\rightarrow  \infty}(-D_p(t))
    \end{equation}

    Proceeding as in section \ref{subsec Hp_0} \cite[see 4.1 for more details]{margalef_concordance} we obtain that the event horizon is:

    \begin{equation}\label{H_e integral}
        H_e=\frac{1}{H_0}\int_{-1}^0\frac{dy}{\sqrt{\sum\Omega_{i0}(1+y)^{n_i}}}
    \end{equation}

  \subsection{At any cosmic time}
    Analogously as in section \ref{subsec Hp(z)}, we can obtain that the event horizon at any $z$ is:

    \begin{equation}\label{H_e(z) final}
        H_e(z)=\frac{1}{1+z}\int_{-1}^z\frac{dx}{H(x)}
    \end{equation}

    If we now take $f(x)=H(x)$ and $g(x) =(1+x)^{m/2}$, then its ratio tends to $H_0\sqrt{\Omega_{N0}}>0$ when $x\rightarrow -1$, so the integrals of $1$ over these functions behave in the same way. Now, applying the $p$-convergence \cite{ponnusamy2011foundations}, we have that for any $z>-1$, the integral of $1/g$ converges if and only if $m/2<1$. Then now we have the following characterization (where again, the limit cases $z=-1$ and $z=\infty$ will be studied in section \ref{sect values}):\vspace*{2ex}

    \centerline{\fbox{$H_e(z)$ exists for $z>-1$ if and only if $m<2$}}
    \mbox{}

    \begin{remark}\mbox{}\\
      Notice that if we consider just one state equation, then $N=m$. Therefore we recover the well known result that there cannot exist both horizons at the same time, and in fact, if $N=m=2$ (state equation of curvature, see remark \refconchap{remark chap curvature}) there exists none of them.
    \end{remark}

\section{Evolution of the Horizons in a General Universe}
    In this section we obtain a general formula for the evolution of the horizons whenever they are defined. In order to do that we just need the following results:
    \begin{align}
      &F(x)=\int_a^{b(x)}f(t)dt \quad\longrightarrow\quad \frac{dF}{dx}=f\left(\rule{0ex}{2.5ex}b(x)\right)b'(x)\\
      &\frac{dz}{dt}=-\frac{a_0}{a^2}\frac{da}{dt}=-\frac{a_0}{a}H=-(1+z)H(z)\label{dz/dt}
    \end{align}

    Where the first expression follows from the well known fundamental theorem of calculus \cite[sect.7.1]{ponnusamy2011foundations} where $f$ is a continuous function defined in a close interval, and the last one is obtained applying first equation \eqref{def redshift}, then equation \eqref{def H(z)} and finally equation \eqref{def redshift} again. Now deriving equations \eqref{H_p(z) final} and \eqref{H_e(z) final}:
    \begin{align*}
       \frac{dH_p}{dz}&=-\frac{1}{(1+z)^2}\int_z^\infty\frac{dx}{H(x)}-\frac{1}{(1+z)H(z)}=\\
                      &=-\frac{H_p}{1+z}-\frac{1}{(1+z)H(z)}\\[3ex]
       \frac{dH_e}{dz}&=-\frac{1}{(1+z)^2}\int_{-1}^z\frac{dx}{H(x)}+\frac{1}{(1+z)H(z)}=\\
                      &=-\frac{H_e}{1+z}+\frac{1}{(1+z)H(z)}
    \end{align*}

    Now, taking into account the chain rule and equation \eqref{dz/dt}, we obtain:

    \begin{equation}\label{velocidades horizontes}
        \boxed{\frac{dH_p}{dt}=H_p(z)H(z)+1}\qquad\qquad\qquad \boxed{\frac{dH_e}{dt}=H_e(z)H(z)-1}
    \end{equation}

    Note that $H_p(z)H(z)$ and $H_e(z)H(z)$ represent, respectively, the recession velocities of the particle and event horizons \eqref{def recession}. Its physical meaning is explained in subsection \ref{sub conclusions}, where all the conclusions are provided.

\section{Relevant Values of the Horizons and Their Velocities}\label{sect values}
\subsection{Horizons at the origin}
  In this section we make a deep study of the behaviour of the particle and event horizons, as well as its derivatives at the origin of the universe $z=\infty$. From now on, when dealing with the particle horizon, we may assume that it is defined for $z>-1$ i.e. that $N>2$. Similarly, when dealing with the event horizon we may assume that $m<2$.
  \begin{description}
    \item[\hspace*{-0.5ex}\fbox{$H_p$}] As $H_p$ exists, then the integral on equation \eqref{H_p(z) final} is convergent for any $z>-1$,  so if we take the limit $z\rightarrow\infty$, $H_p$ goes to zero. Then
      \begin{equation}\label{h_p(inf)}
        \lim_{z\rightarrow \infty}H_p(z)=0
      \end{equation}
    \item[\hspace*{-0.39ex}\fbox{$H_e$}] If we try to compute the limit of $H_e(z)$ when $z\rightarrow\infty$, we obtain zero if $N>2$ (as in this case, the integral appearing in the expression of $H_e$ is convergent), and an indetermination $\frac{\infty}{\infty}$ if $N\leq2$. For the latter case, applying L'H{\^o}pital's rule and the fundamental theorem of calculus, we obtain:
    \begin{equation}\label{h_e(inf)}
        \lim_{z\rightarrow\infty}H_e(z)=\lim_{z\rightarrow\infty}\frac{1}{H(z)}=\left\{\begin{array}{ll}
            0                                & \text{\ if\ } N>0\\[0.8ex]
            \dfrac{1}{H_0\sqrt{\Omega_{N0}}} & \text{\ if\ } N=0\\[2ex]
            \infty                           & \text{\ if\ } N< 0
        \end{array}\right.
    \end{equation}
  \end{description}

  Once we have computed the limits of the horizons themselves, let us deal with their velocities i.e. how the proper distance of the horizon varies with respect to the time coordinate $t$. In order to do that, first notice that:
  \begin{align}
     &\lim_{z\rightarrow\infty}\frac{H(z)}{1+z}=\left\{\begin{array}{ll}
            \infty               & \text{\ if\ } N>2\\[1.3ex]
            H_0\sqrt{\Omega_{N0}} & \text{\ if\ } N=2\\[2ex]
            0                    & \text{\ if\ } N<2
        \end{array}\right.\label{H(z) z->inf}\\[2ex]
     & H'(z)=\frac{H_0^2}{2H}\sum n_i\Omega_{i0}(1+z)^{n_i-1}
  \end{align}

  Where the last formula is valid thanks to the hypothesis we assume on $H$ (see remarks after eq. \eqref{H(z) sumatorio}).

  \begin{description}
    \item[\hspace*{-1.8ex}\fbox{$\frac{dH_p}{dt}$}] As we are assuming $N>2$, if we take the limit $z\rightarrow\infty$ of $H_pH$ we obtain an indetermination $0\cdot\infty$:
        \begin{align*}
          \lim_{z\rightarrow\infty} H_pH&=\lim_{z\rightarrow\infty}\frac{1}{(1+z)/H(z)}\int_z^\infty \frac{dx}{H(x)}\overset{L'H}{=}\\[2ex]
                                        &=\lim_{z\rightarrow\infty}\frac{-1/H(z)}{\frac{H-(1+z)H'}{H^2}}=\lim_{z\rightarrow \infty}\frac{-H}{H-(1+z)H'}=\\[1.5ex]
                                        &=\left(\lim_{z\rightarrow\infty}\frac{(1+z)H'}{H}-1\right)^{-1}=\\[1.5ex]
                                        &=\left(\frac{1}{2}\lim_{z\rightarrow\infty}\frac{\sum n_i\Omega_{i0}(1+z)^{n_i}}{\sum\Omega_{i0}(1+z)^{n_i}}-1\right)^{-1}=\\[1.5ex]
                                        &=\left(\frac{N}{2}-1\right)^{-1}=\frac{2}{N-2}
        \end{align*}

        Then, from equation \eqref{velocidades horizontes} we have

        \begin{equation}
          \lim_{t\rightarrow 0}\frac{dH_p}{dt}=\frac{N}{N-2}
        \end{equation}
    \item[\hspace*{-1.8ex}\fbox{$\frac{dH_e}{dt}$}] We are now assuming that $m<2$, but when $z$ tends to infinity, the dominant term is the one with the largest $n_i$ i.e. the one associated to $N$, and so we have to consider three possibilities:\vspace*{1.5ex}
        \begin{itemize}
          \item[$N>2$] In that case,
              \[\lim_{z\rightarrow\infty}\int_{-1}^z \frac{dx}{H(x)}<\infty\]
              and equation \eqref{H(z) z->inf} implies that velocity of the event horizon tends to infinite.
          \item[$N=2$] Now the previous limit is divergent, but again equation \eqref{H(z) z->inf} implies that velocity of the event horizon tends to infinite.
          \item[$N<2$] In this last case we have, as in the previous one, that the integral is divergent, but now notice that equations \eqref{velocidades horizontes} and \eqref{H(z) z->inf} lead to an indetermination $0\cdot\infty$. Using exactly the same computations as in the $dH_p/dt$ case (but now, when applying the fundamental theorem of Calculus no minus sign appears) we have:

              \begin{equation}
                \lim_{t\rightarrow 0}\frac{dH_e}{dt}=\frac{N}{2-N}
              \end{equation}
        \end{itemize}
  \end{description}

  In table \ref{table relevant values} we have summarized all these data.

\subsection{Horizons at the far future}
 In this section we make a deep study of the behaviour of the particle and event horizons, as well as its derivatives at the far future $z=-1$.

  \begin{description}
    \item[\hspace*{-0.5ex}\fbox{$H_p$}] Notice that the integrand on equation \eqref{H_p(z) final} is positive everywhere (due to the expansion of the universe, see explanation below eq. \eqref{def H(z)}), and it can only vanish for the value $x=z=-1$, then the integral exists (though possible infinite) and it is greater than zero, so the fraction multiplying the integral in \eqref{H_p(z) final} leads to

        \begin{equation}\label{h_p(-1)}
          \lim_{z\rightarrow -1}H_p(z)=\infty
        \end{equation}

    \item[\hspace*{-0.39ex}\fbox{$H_e$}] If we try to compute the limit of $H_e$, and indetermination $\frac{0}{0}$ appears. Proceeding in analogy with the previous section, we obtain:

        \begin{equation}\label{h_e(-1)}
          \lim_{z\rightarrow  -1}H_e(z)=\lim_{z\rightarrow -1}\frac{1}{H(z)}=\left\{\begin{array}{ll}
              0                                & \text{\ if\ } m<0\\[0.8ex]
              \dfrac{1}{H_0\sqrt{\Omega_{m0}}} & \text{\ if\ } m=0\\[2ex]
              \infty                           & \text{\ if\ } m>0
          \end{array}\right.
        \end{equation}

  \end{description}

  Now, in order to compute the limits of both velocities, notice that
  \begin{align}
     &\lim_{z\rightarrow -1}\frac{H(z)}{1+z}=\left\{\begin{array}{ll}
            \infty                & \text{\ if\ } m<2\\[1.3ex]
            H_0\sqrt{\Omega_{m0}} & \text{\ if\ } m=2\\[2ex]
            0                     & \text{\ if\ } m>2
        \end{array}\right.
  \end{align}

  \begin{description}
    \item[\hspace*{-1.8ex}\fbox{$\frac{dH_p}{dt}$}] We are assuming $N>2$, but when $z$ tends to $-1$, the dominant term is the lowest one. Making the same study as in the $dH_e/dt$ case for $z\rightarrow\infty$ we obtain:
        \[\lim_{t\rightarrow  \infty}\frac{dH_p}{dt}=\left\{\begin{array}{ll}
            \infty         & \text{\ if\ } m\leq 2\\[2ex]
            \dfrac{m}{m-2} & \text{\ if\ } m>2
        \end{array}\right.\]
    \item[\hspace*{-1.8ex}\fbox{$\frac{dH_e}{dt}$}] We are now assuming that $m<2$, and if we now proceed as in the $dH_p/dt$ case for $z\rightarrow\infty$ (as an $\infty\cdot 0$ indetermination arises) we have:

        \begin{equation}
          \lim_{t\rightarrow\infty}\frac{dH_e}{dt}=\frac{m}{2-m}
        \end{equation}

  \end{description}

  In table \ref{table relevant values} we have summarized all these data.\vspace*{2ex}

 \begin{table}[ht!]
      \centering
      \begin{tabular}{c|c c |c c c c}
      \hline
      \rule{0ex}{4.5ex}                     &    $\begin{array}{c}  H_p(z)\\[1ex] (N>2) \end{array}$ & $\begin{array}{c}  H_e(z)\\[1ex] (m<2) \end{array}$ & $\dfrac{dH_p}{dt}\quad (N>2)$  & $\dfrac{dH_e}{dt}\quad (m<2)$\\[2.2ex]
      \hline\hline
      \rule{0ex}{8ex}\begin{tabular}{c}
        Origin of\\ the universe\\ $t=0$
      \end{tabular}&  $0$        & $\left\{\begin{array}{ll}
           									    	  0                                  & \text{\ if\ } N>0\\[0.8ex]
            										  \dfrac{1}{H_0\sqrt{\Omega_{N0}}}   & \text{\ if\ } N=0\\[2ex]
            										  \infty                             & \text{\ if\ } N< 0
        									       \end{array}\right.$   & $\dfrac{N}{N-2}$     & $\left\{\begin{array}{ll}
                                                                                                 \infty         & \text{\ if\ } N\geq 2\\[1.5ex]
                                                                                                 \dfrac{N}{2-N} & \text{\ if\ } N<2
                                                                                               \end{array}\right.$\\[7ex]
      \begin{tabular}{c}Far Future\\ $t=\infty$\end{tabular}    &   $\infty$     & $\left\{\begin{array}{ll}
            										  0                                  & \text{\ if\ } m<0\\[0.8ex]
            										  \dfrac{1}{H_0\sqrt{\Omega_{m0}}}   & \text{\ if\ } m=0\\[2ex]
            									          \infty                             & \text{\ if\ } m> 0
        									       \end{array}\right.$   & $\left\{\begin{array}{ll}
                                                                                \infty         & \text{\ if\ } m\leq 2\\[1.5ex]
                                                                                \dfrac{m}{m-2} & \text{\ if\ } m>2
                                                                            \end{array}\right.$ & $\dfrac{m}{2-m}$     \\[6ex]
      \hline
      \end{tabular}
      \centering\caption{Some important values of the horizons and their velocities in geometrical units.}\label{table relevant values}
    \end{table}

    \vspace*{2ex}

  \begin{remarks}\mbox{}\renewcommand{\labelenumi}{\thesection.\arabic{enumi}}
    \begin{enumerate}
      \setcounter{enumi}{\value{theorem}}
      \item Whenever it exists, $H_p$ always comes from zero at the origin of the universe, and tends to infinite at the far future no matter whatsoever which and how many state equations are considered.
      \item Notice that $H_e$ is constant for every $z\in(-1,\infty)$ if and only if just one state equation with $N=m=0$ is considered, in which case, $H_e(z)=H_0\sqrt{\Omega_0}$ for every $z$.
      \item Though physically it is not accepted in the current theories, we can deduce that if $m<0$ or equivalently $w_m<-1$, we would have that the event horizon speed becomes strictly negative for some $z>-1$ as in the limit it is strictly negative. We proved in \cite{margalef_concordance} that for the concordance Universe (two state equations with $m=0$ and $N=3$) such a thing does not happen i.e. it always remains positive.
      \item Notice that all the results we have obtained so far, depend on the state equations associated to $N$ or $m$, and some of them do not depend on the state equations considered.
      \setcounter{theorem}{\value{enumi}}
    \end{enumerate}
  \end{remarks}

  \subsection[When are Hp and He equal?]{When are $H_p$ and $H_e$ equal?}
    It is interesting to know whether both horizons might have or not the same values, and if so, how often this could happen. In order to do that, let us look for a $z_0$ such that $H_p(z_0)=H_e(z_0)$, where we are obviously assuming that $N>2$ and $m<2$ so both horizons exist. First notice that if $z_0=-1$, then according to equation \eqref{h_p(-1)}, they have both to be infinite and in that case we know from equation \eqref{h_e(-1)} that $m>0$.\vspace*{2ex}

    The previous solution has little interest, so let us now look for other possible solutions such that $z_0>-1$. According to equations \eqref{H_p(z) final} and \eqref{H_e(z) final}, and cancelling out the fractions before the integrals:
    \[\int_{z_0}^\infty\frac{dx}{H(x)}=\int_{-1}^{z_0}\frac{dx}{H(x)}\]
    The fact that $N>2$ and $m<2$ ensures that the integral from $z=-1$ to infinite is convergent, then:
    \begin{align*}
       L&\equiv\int_{-1}^\infty\frac{dx}{H(x)}=\int_{-1}^{z_0}\frac{dx}{H(x)}+\int_{z_0}^\infty\frac{dx}{H(x)}=\\[1.5ex]
        &=(1+z_0)\biggl(H_e(z_0)+H_p(z_0)\biggr)\overset{\star}{=}\\
        &=(1+z_0)2H_e(z_0)=2\int_{-1}^{z_0}\frac{dx}{H(x)}
    \end{align*}

    Where in the $\star$ equality we have used that $H_p(z_0)=H_e(z_0)$. Then finally:

    \begin{equation}\label{last_equality}
        \int_{-1}^{z_0}\frac{dx}{H(x)}=\frac{L}{2}
    \end{equation}

    As $H(x)$ is positive wherever it is defined, the integral of the left hand side of the equation is strictly increasing with respect to $z_0$, and it goes from $0$ when $z_0=-1$, and tends to $L$ when $z_0\rightarrow\infty$. Then there exists a $z_0\in(-1,\infty)$ such that verifies the previous equation \eqref{last_equality} and then verifies the equation $H_p(z_0)=H_e(z_0)$. If we suppose that there exist $z_1$ and $z_2$ verifying the previous equation, then:
    \begin{align*}
       \int_{z_1}^{z_2}\frac{dx}{H(x)}&=\int_{z_1}^{-1}\frac{dx}{H(x)}+\int_{-1}^{z_2}\frac{dx}{H(x)}=\\[2ex]
                                      &=\int_{-1}^{z_2}\frac{dx}{H(x)}-\int_{-1}^{z_1}\frac{dx}{H(x)}=\frac{L}{2}-\frac{L}{2}=0
    \end{align*}
        and again, as $H$ is positive, we have that $z_1=z_2$. Finally, taking into account the following chain of equivalences:
    \begin{align*}
       H_e(z)>H_p(z)\quad&\Longleftrightarrow\quad \int_{-1}^z\frac{dx}{H(x)}>\int_z^{\infty}\frac{dx}{H(x)} \quad\Longleftrightarrow\quad 2\int_{-1}^z \frac{dx}{H(x)}>\int_{-1}^\infty \frac{dx}{H(x)}=L \quad\Longleftrightarrow\\[2.5ex]
       \quad&\Longleftrightarrow\quad \int_{-1}^z\frac{dx}{H(x)}>\frac{L}{2}=\int_{-1}^{z_0}\frac{dx}{H(x)}\quad\Longleftrightarrow\quad \int_{z_0}^z\frac{dx}{H(x)}>0\quad\Longleftrightarrow\quad z>z_0
    \end{align*}

    where in the last equivalence we have use once again that $H$ is positive, we may summarize all the previous computations in the following result:

    \begin{theorem}\mbox{}\\
       There exists one and only one $z_0>-1$ such that
       \[\left|\begin{array}{ll}
            H_e(z)>H_p(z) & \text{for every }z\in(z_0,\infty)\\[2ex]
            H_e(z_0)=H_p(z_0) \\[2ex]
            H_e(z)<H_p(z) & \text{for every }z\in(-1,z_0)
       \end{array}\right.\]
    \end{theorem}

    \begin{remark}\mbox{}\\
       Obviously, the numerical value of the $z_0$ itself depends on which state equations we take into account and also on the numerical value of all the $\Omega_{i0}$.
    \end{remark}

\section{Results and Interpretation}
  \subsection{Application to the concordance universe}
    In this section we are going to prove that in fact we are generalizing the previous results obtained for a concordance universe in \cite{margalef_concordance} i.e. taking into account two state equations (dust and cosmological constant). For these two state equations we have that $N=3$ and $m=0$, then table \ref{table relevant values}, recovering the constant $c$, becomes table \ref{table2}:

    \begin{table}[ht!]
      \centering
      \begin{tabular}{l|c c |c c c c}
        \hline
        \rule{0ex}{4.5ex}                     &   $z$  &   $t$  &$H_p(z)$&$H_e(z)$ & $\dfrac{dH_p}{dt}$ & $\dfrac{dH_e}{dt}$\\[2ex]
        \hline\hline
        \rule{0ex}{3ex}Origin of the universe &$\infty$&   $0$  &   $0$  &   $0$   &         $3c$        &    $\infty$ \\[1.5ex]
                     Future                   &  $-1$  &$\infty$&$\infty$&$\dfrac{c}{H_0\sqrt{\Omega_{m0}}}$ & $\infty$ & $0$           \\[2ex]
        \hline
      \end{tabular}\caption{Relevant values for the concordance universe.}\label{table2}
    \end{table}

    which agrees completely with table $1$ of \cite{margalef_concordance}, where we considered the concordance universe  i.e. euclidean, with cosmological constant and dust.\vspace*{2ex}

\subsection{Application to the concordance universe with radiation}

    We can go further in our analysis as it is widely accepted that for the origin of the universe, the state equation that dominated was radiation (then we have $N=4$ instead of $N=3$). In this case it holds that all the data are the same as for the concordance universe, but for the velocity of the particle horizon at the origin (see table \ref{table3}):
    \[\lim_{t\rightarrow0}\frac{dH_p}{dt}=2c\]

   \begin{table}[ht!]
      \centering
      \begin{tabular}{l|c c |c c c c}
        \hline
        \rule{0ex}{4.5ex}                     &   $z$  &   $t$  &$H_p(z)$&$H_e(z)$ & $\dfrac{dH_p}{dt}$ & $\dfrac{dH_e}{dt}$\\[2ex]
        \hline\hline
        \rule{0ex}{3ex}Origin of the universe &$\infty$&   $0$  &  $0$   &   $0$   &          $2c$       &    $\infty$ \\[1.5ex]
                     Future                   &  $-1$  &$\infty$&$\infty$&$\dfrac{c}{H_0\sqrt{\Omega_{m0}}}$ & $\infty$ & $0$           \\[2ex]
        \hline
      \end{tabular}\caption{Relevant values for the concordance universe with radiation.}\label{table3}
    \end{table}

  \subsection{Conclusions}\label{sub conclusions}
    In this paper, we have first obtained integral expressions for the particle and the event horizons considering the extremely general case of a universe with or without curvature ruled by countably infinitely many state equations (under some reasonable hypotheses on those state equations that are fulfilled when only finitely many state equations, no matter which one, are considered). After some suitable manipulation, we have computed their first derivative, obtaining these extremely simple equations:
    \[\frac{dH_p}{dt}=H_p(z)H(z)+1\qquad \qquad \frac{dH_e}{dt}=H_e(z)H(z)-1\]
    where $H_pH$ and $H_eH$ are the recession velocities of the particle and event horizons respectively. These expressions generalize the previous stated results derived for just one constant state equation and null curvature \cite{harrison} and for two state equation \cite{margalef_concordance}. Notice that all these results and values are independent of the current values of the involved constants.\vspace*{2ex}

    The equations of the velocity of the horizon have a remarkable physical meaning. As the recession velocity $H_pH$ is the instantaneous velocity of an object located at the distance of the particle horizon $H_p$, from the first equation we deduce that the instantaneous velocity of the surface of the horizon particle is faster than the one of the objects over the particle horizon, then more and more objects are entering into the particle horizon and they will never get out. Analogously, the second equation stands that the recession speed of the event horizon is slower than that of the objects over the event horizon, and then more and more objects are disappearing from the event horizon and they will never get into again.\vspace*{2ex}

    We have also obtained the value of the horizons and their velocities at origin of the universe and for the far future with the great generality stated above. As showed in table \ref{table relevant values}, some of them depend on which state equations are considered. We have obtained also the expected result, that these values only depend on the respective dominant state equation. Finally we have obtained that there exists one and only one $z_0>-1$ where both horizons have the same value, and also that for lower values of the redshift $z$ it holds $H_e(z)<H_p(z)$ while for larger values $H_e(z)>H_p(z)$.\vspace*{2ex}

    \paragraph{Acknowledgements}\mbox{}\\
    This work was partially supported by the Spanish Ministry of Economy and Competitiveness (MINECO) under the grant AYA2011-29517-C03-01.

  \hypersetup{urlcolor=blue}


\begin{thebibliography}{99}
    \bibitem{cepa2007cosmologia} J. Cepa, \emph{Cosmolog\'ia F\'isica}, \href{http://www.akal.com/libros/CosmologIa-FIsica/9788446025337}{Akal}, Madrid, Spain (2007)
    \bibitem{harrison}E. Harrison, \emph{Hubble Spheres and Particle Horizons}, \href{http://adsabs.harvard.edu/full/1991ApJ...383...60H}{\emph{Astrophys. J.}} \textbf{383} (1991) 60.
    \bibitem{kolb1990early} E.W. Kolb and M.S. Turner, \emph{The early universe}, \href{http://www.westviewpress.com/book.php?isbn=9780201626742}{West View Press} (1990).
    \bibitem{margalef_concordance} B. Margalef--Bentabol, J. Margalef--Bentabol and J. Cepa, \emph{Evolution of the cosmological horizons in a concordance universe}, \href{http://iopscience.iop.org/1475-7516/2012/12/035}{\emph{JCAP}} \textbf{12} (2012) 035 [\href{http://arxiv.org/abs/1302.1609}{arXiv:1302.1609}]
     \bibitem{ponnusamy2011foundations} S. Ponnusamy, \emph{Foundations of mathematical analysis}, \href{http://www.springer.com/birkhauser/mathematics/book/978-0-8176-8291-0}{Birkh\"{a}user Basel}, Berlin, Germany (2011).
    \bibitem{Rindler}W. Rindler, \emph{Visual Horizons in World-Models, \href{http://articles.adsabs.harvard.edu//full/1956MNRAS.116..662R/0000662.000.html}{Mon. Not. Roy. Astr. Soc.}} \textbf{116} (1956) 662.
\end{thebibliography}
\end{document}